\documentclass[journal=jacsat,manuscript=article,layout=twocolumn]{achemso}

\usepackage[version=3]{mhchem} 
 \SectionNumbersOn
\usepackage{graphicx}

\usepackage{bm}
\usepackage{braket}
\usepackage{amsfonts}
\usepackage{amsmath}
\usepackage{amssymb}
\usepackage{cleveref}
\usepackage[varg]{txfonts}
\usepackage[font=footnotesize,labelfont=bf]{caption}
\usepackage{soul}
\usepackage[small]{titlesec}

\usepackage{xcolor}
\usepackage[section]{placeins}

\usepackage{booktabs}

\usepackage{float}
\author{Jannis Kockl{\"a}uner}
\affiliation{Faculty of Chemistry and Food Chemistry, Technische Universit\"at Dresden, 01062 Dresden, Germany}
\email{jannis.kocklaeuner@tu-dresden.de}

\author{Majid Shaker}
\affiliation{Chair of Physical Chemistry II, Friedrich-Alexander-Universität Erlangen-N\"urnberg, Egerlandstr. 3, 91058 Erlangen, Germany}

\author{Maximilian Muth}
\affiliation{Chair of Physical Chemistry II, Friedrich-Alexander-Universität Erlangen-N\"urnberg, Egerlandstr. 3, 91058 Erlangen, Germany}

\author{Simon Steinbach}
\affiliation{Chair of Physical Chemistry II, Friedrich-Alexander-Universität Erlangen-N\"urnberg, Egerlandstr. 3, 91058 Erlangen, Germany}

\author{Christoph Oleszak}
\affiliation{Chair of Organic Chemistry II, Friedrich-Alexander-Universität Erlangen-N\"urnberg, Nikolaus-Fiebiger-Str. 10, 91058 Erlangen, Germany}

\author{Ole Lytken}
\affiliation{Chair of Physical Chemistry II, Friedrich-Alexander-Universität Erlangen-N\"urnberg, Egerlandstr. 3, 91058 Erlangen, Germany}

\author{Hans-Peter Steinr\"uck}
\affiliation{Chair of Physical Chemistry II, Friedrich-Alexander-Universität Erlangen-N\"urnberg, Egerlandstr. 3, 91058 Erlangen, Germany}

\author{Dorothea Golze}
\affiliation{Faculty of Chemistry and Food Chemistry, Technische Universit\"at Dresden, 01062 Dresden, Germany}
\email{dorothea.golze@tu-dresden.de}

\title[gwc]{Decoding Shake-up Satellites in XPS through Large-Scale ab initio Simulations: Spectral Signatures of Ring Fusion in Porphyrins}

\abbreviations{GW,HPC,FHI-aims}

\let\oldmaketitle\maketitle
\let\maketitle\relax
\begin{document}
\linespread{1.1}
\fontsize{10}{12}\selectfont
\twocolumn[
  \begin{@twocolumnfalse}
    \oldmaketitle
    \begin{abstract}
\fontsize{10}{12}\selectfont

In X-ray photoelectron spectroscopy (XPS), shake-up satellites arise when core ionization is accompanied by simultaneous charge-neutral valence excitations. Although these satellites can contain detailed structural information, they are rarely interpreted due to the lack of accurate and scalable theoretical methods. Here, we develop and apply a many-body perturbation theory framework within the $GW$ plus cumulant ($GW+C$) approach that enables accurate predictions of shake-up satellites in large molecular systems. For unfused, mono-fused, and doubly fused porphyrin derivatives with up to 170 atoms, we achieve excellent agreement with experiment, reproducing both main photoionization signals and satellite features within $0.2-0.3$~eV. We show that ring fusion strongly affects satellite features, whereas the N~1s photoionization signals remain unchanged. Our calculations reveal the mechanism behind these changes, identifying the spatial localization of valence excitations as the driving force. This work not only deepens understanding of the shake-up mechanism in porphyrins but also shows how predictive computations can unlock the chemical information encoded in satellites.

\end{abstract}
  \end{@twocolumnfalse}
  ]

\begin{tocentry}
\centering
\includegraphics[width=0.89\columnwidth]{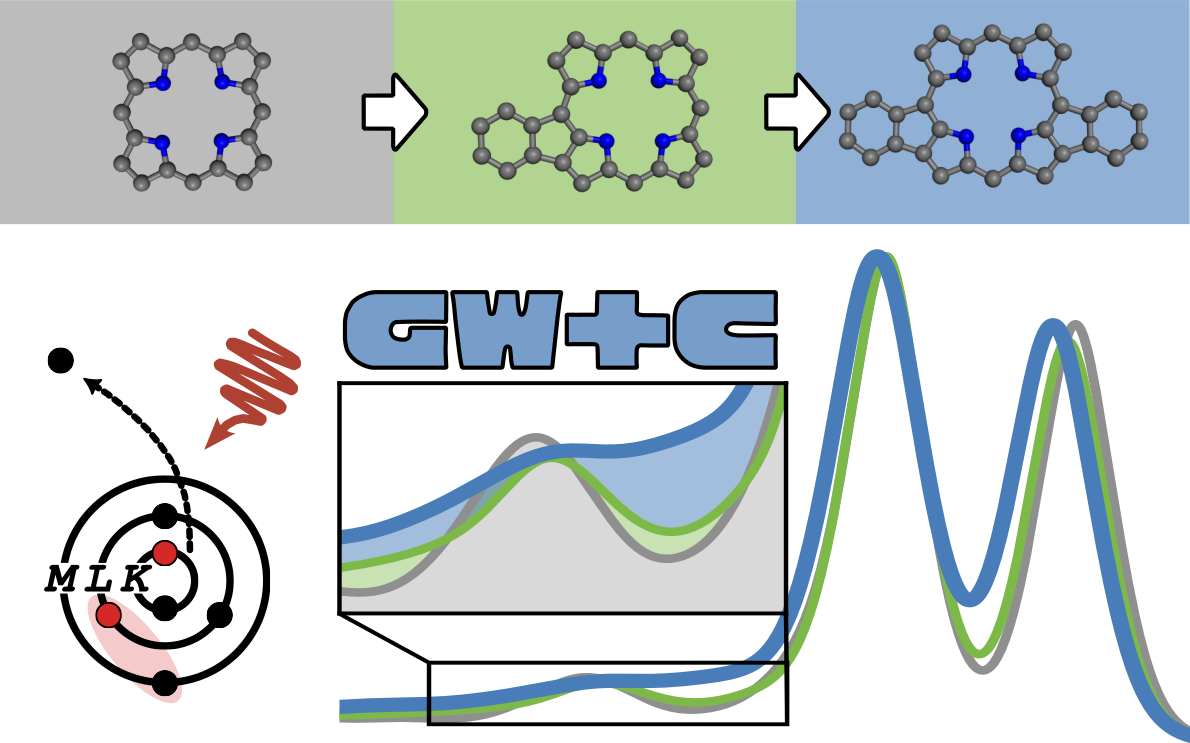}
\end{tocentry}

\section{Introduction}
X-ray photoelectron spectroscopy (XPS) is a key experimental technique for extracting chemical and structural information across a wide range of materials. In XPS, core electrons are ejected upon irradiation with high-energy X-rays, and the measured binding energies serve as element-specific fingerprints that are highly sensitive to the local chemical environment.
\cite{de2008core,bagus2018extracting, van2011x} 
\par
Beyond primary ionization peaks, XP spectra often exhibit shake-up satellites at higher binding energies, which originate from charge-neutral valence excitations coupled to the core-ionization process.\cite{aaberg1967theory, aagren1992core, bagus2013interpretation} These satellite features are especially prominent in extended systems, transition metal complexes, and large $\pi$-conjugated compounds.\cite{de2008core, scholl2004line, gengenbach2021practical, bagus2013interpretation}
While satellites complicate the spectral interpretation and peak assignments,\cite{gengenbach2021practical, major2020practical} they also provide an opportunity to probe the coupling between localized core states and valence-shell electronic excitations. 
Consequently, shake-up satellites can provide additional chemical information on oxidation\cite{otamiri1990ammoxidation, velasquez2005chemical, meda2004decomposition, biesinger2011resolving, de2001high} and spin states,\cite{yin1974paramagnetism, grosvenor2004investigation, tanuma2023noncontact, de2001high} conformational changes,\cite{wang2024configuration, mosaferi2024fingerprint} and chemical bonding.\cite{lombana2025competing, li2024core}.
\par
Still, shake-up satellites are analyzed much less frequently than main ionization peaks, mainly because their lower intensity increases the effort required for measurement and the lack of accurate, scalable theoretical tools often makes a faithful peak assignment impossible.
First-principles methods like $\Delta$-self-consistent-field ($\Delta$SCF) methods based on density-functional theory (DFT),\cite{Susi2015,vines2018prediction, klein2021nuts,kahk2019accurate,kahk2021core,besley2021modeling, kahk2022predicting} as well as many-body perturbation theory in the $GW$ approximation, \cite{golze2018core, golze2020accurate, keller2020relativistic, li2022benchmark, golze2022accurate, mejia2022basis, galleni2022modeling, galleni2024c1s, zarrouk2024experiment, bruneval2024fully,bruneval2025gw2sosex} are established as reliable tools for the interpretation of main ionization peaks. They provide absolute core-level binding energies within 0.2-0.3~eV of the experiment, \cite{ li2022benchmark, kahk2019accurate} but they systematically fail to capture satellite features: $\Delta$SCF does not account for them at all, and $GW$ systematically predicts incorrect satellite positions and intensities due to missing vertex corrections.\cite{langreth1970singularities, martin_interacting_2016}
\par
For finite systems, wave-function-based approaches (e.g., configuration interaction or coupled-cluster theory) can describe shake-up states accurately but are computationally prohibitive beyond small molecules.\cite{kuramoto2005theoretical, ehara2005sac,kuramoto2005c, sankari2006high, chatterjee2019second, marie2024reference, rehr2020equation,vila2020real, vila2021equation, vila2022real, pathak2023real} Alternative methods like the equivalent core-hole (ECH) approximation combined with time-dependent DFT (TDDFT) offer improved efficiency,\cite{brena2005functional,brena2006time, deng2014comparative, li2024core, wang2024configuration, lombana2025competing} but suffer from severe limitations, including incorrect state ordering,\cite{lee1994identification} poor performance for shallow or semicore levels,\cite{plashkevych2000validity} and the inability to capture physical effects like multiplet splitting, lifetime broadening, and correlation effects.\cite{bagus2020limitations, olovsson2009all}
Furthermore, simulating shake-up states in periodic systems with ECH theory involves a charged supercell, prohibiting a seamless transition to extended systems.
\par
In the solid state, advanced many-body techniques have been developed to describe satellite features arising from collective plasmon or charge-transfer excitations, rather than localized molecular shake-up satellites.
For instance, combining density functional theory with dynamical mean-field theory (LDA+DMFT) accurately captures charge-transfer satellites in strongly correlated transition-metal oxides such as SrTiO$_3$ and TiO$_2$.\cite{hariki2017lda, ghiasi2019charge, hariki2022satellites}
Likewise, the $GW$ approach combined with the cumulant expansion ($GW+C$) has been widely applied to plasmon satellites in various materials.\cite{aryasetiawan1996multiple, zhou2015dynamical, vigil2016dispersion, mcclain2016spectral, zhou2018cumulant, kas2016particle,guzzo2011valence, kheifets2003spectral, lischner2015satellite, caruso2015band, gumhalter2016combined, nakamura2016ab,regoutz2019insights,lischner2013physical, guzzo2014multiple}
However, these methods have not been applied to core-level shake-up satellites in molecular systems.\par
To address this challenge, we recently introduced a fully \textit{ab initio} many-body framework for simulating core-level shake-up satellites based on the $GW+C$ approach.\cite{kocklauner2025gw} Although applied so far only to molecules, the method is conceptually suited for solids as well. We developed a numerically efficient implementation that reproduces both main peaks and satellites with excellent accuracy and treats them consistently on the same footing, as demonstrated for the C1s shake-up spectra of the acene series up to pentacene.\cite{kocklauner2025gw} Here, we extend this predictive framework to molecules with up to 200 atoms - an unprecedented system size for accurate satellite calculations - enabling the quantitative interpretation of complex XP spectra in chemically relevant systems such as porphyrins and their derivatives.
\par
Porphyrins are a widely studied class of macrocyclic compounds with a $\pi$-conjugated macrocycle and four nitrogen atoms that can coordinate metal ions. Due to their rich optical and electronic properties, porphyrins play a central role as building blocks for optoelectronic devices, organic photovoltaics and catalysis.\cite{tanaka2015conjugated, wolf2017rigid, jurow2010porphyrins, park2020applications}
Extending the conjugated $\pi$-system through additional ring fusion has proven an effective strategy to tune absorption energies, charge transport, and excited-state lifetimes.\cite{davis2011porphyrin,gill2004facile,he2017fusing, mateo2020surface, martin2020oxidative, oleszak2024fused}
\par
In this work, we present a combined experimental and theoretical study examining the impact of ring fusion on the main and satellite features in the N~1s XP spectra of porphyrin derivatives.
We demonstrate that our $GW+C$ approach can reliably predict the full XP spectra and precisely identify the specific excitations that give rise to shake-up satellites. Our advanced theoretical framework provides a detailed understanding of the shake-up mechanism in porphyrins, clarifies its relation to molecular structure, explains the origins of the observed changes, and shows how the scope of standard XPS can be broadened when satellites are reliably interpreted.
\par
In the following, we briefly introduce the $GW+C$ theory (Section~\ref{sec:methods}), describe the porphyrin systems and methodology (Section~\ref{sec:systems_and_methods}), present and analyze the results (Section~\ref{sec:results}), and close with a summary and outlook (Section~\ref{sec:conclusion}).
\section{Methods}
\label{sec:methods}
\subsection{Theoretical core-level photoemission spectroscopy}
\label{sec:gw_theory}
Photoemission spectroscopy techniques, such as XPS, measure the photocurrent $J(\omega)$. Within the sudden approximation,\cite{hedin1999correlation,onida_electronic_2002, martin_interacting_2016} the spectral function $A(\omega, \mathbf{r}, \mathbf{r}^\prime)$ of an interacting electron system can be directly related to $J(\omega)$.
The spectral function $A(\omega, \mathbf{r}, \mathbf{r}^\prime)$ describes the probability of removing or adding an electron at energy $\omega$ and is defined as
\begin{equation}
  A(\omega, \mathbf{r}, \mathbf{r}^\prime) = \frac{1}{\pi} \left| \mathrm{Im} \, G(\omega, \mathbf{r}, \mathbf{r}^\prime) \right|
  \label{eq:spectral_function}
\end{equation}
where $G(\omega, \mathbf{r}, \mathbf{r}^\prime)$ is the one-particle Green's function and describes the propagation of an electron or hole in an interacting many-body system.
The poles of $G$ correspond to the electron removal or addition energies, which appear as main peaks in the spectral function. We also refer to these excitations as “charged excitations” because the system is ionized.
\par
For deep core-states $c$, we can approximate $G$ using a cumulant \textit{ansatz} $G_c^C(t)$ for the time-dependent Green's function\cite{langreth1970singularities,aryasetiawan1996multiple,guzzo2011valence, kas_cumulant_2014, zhou2015dynamical, zhou2018cumulant, vila2020real, loos2024cumulant, kocklauner2025gw}
\begin{equation}
  G_c^C(t) = G_{0,c}(t) \exp\left[C_c(t)\right]
  \label{eq:cumulant_propagator}
\end{equation}
where $C_c(t)$ is a cumulant function introducing many-body correlation effects to a mean-field (DFT) propagator $G_0$.
The resulting frequency-dependent cumulant Green's function takes the form of an infinite series
\begin{equation}
    G^C_{c}(\omega) = Z_c \left[ G^\mathrm{QP}_{c}(\omega) + G^1_{c}(\omega) + \frac{1}{2}G^2_{c}(\omega) + \dots \right]
\end{equation}
with $Z_c$ being a renormalization factor defining the weight of the main ionization peak compared to the satellite spectrum.
\par
Each term in the series corresponds to a distinct excitation: $G^\mathrm{QP}_{c}(\omega)$ yields the photoionization peak or so-called  quasiparticle (QP) peak, while $G^{1}_{c}(\omega)$ and $G^{2}_{c}(\omega)$ generate shake-up satellites from single- and double-excitations, respectively. The satellite intensities decrease in a Poisson-like fashion with excitation order. As shown in our previous work~\cite{kocklauner2025gw}, higher-order terms $G^{2,3,\dots}_{c}(\omega)$ carry vanishingly small weight in molecules. The computed XP spectrum is thus dominated by the QP peak produced by $G^\mathrm{QP}_{c}(\omega)$ and the single-excitation satellites generated by $G^1_{c}(\omega)$. The terms are defined as as~\cite{loos2024cumulant, kocklauner2025gw}
\begin{align}
  G^\mathrm{QP}_{c}(\omega) &= \frac{1}{\omega - \epsilon^\mathrm{QP}_{c} + i\eta}
  \label{eq:qp_propagator} \\
  G^1_{c}(\omega) &= \sum_\nu \frac{\gamma^\nu_c}{\omega - \epsilon^\mathrm{QP}_{c} + \Omega^\nu + i\eta}
  \label{eq:first_order_satellite}
\end{align}
where $\eta$ is an infinitesimal broadening parameter. $\epsilon_c^{\text{QP}}$ denotes the QP energy and $\Omega^\nu$  the charge-neutral excitation energy for the formation of an electron-hole pair (exciton). The first-order satellite intensities $\gamma^\nu$ are given in terms of the transition moments $\rho^\nu_{cc}$ 
\begin{equation}
  \gamma^\nu_c = \left( \frac{\rho^\nu_{cc}}{\Omega^\nu} \right)^2
  \label{eq:gamma_definition}
\end{equation}

The charge-neutral excitation energies $\Omega^{\nu}$ are obtained by solving the following eigenvalue problem
\begin{equation}
  \begin{bmatrix}
    A & B \\
    -B & -A
  \end{bmatrix}
  \begin{bmatrix}
    X^\nu \\
    Y^\nu
  \end{bmatrix}
  = \Omega^\nu
  \begin{bmatrix}
    X^\nu \\
    Y^\nu
  \end{bmatrix}
  \label{eq:rpa_eigenproblem}
\end{equation}
which is a Casida-like equation based on the random phase approximation (RPA). The matrices are defined as
\begin{equation}
  \begin{aligned}
    A_{ia,jb} &= \delta_{ij}\delta_{ab} (\epsilon_a - \epsilon_i) + (ia|jb) \\
    B_{ia,jb} &= (ia|bj)
  \end{aligned}
  \label{eq:rpa_matrices}
\end{equation}
where $\epsilon_{i,a}$ denote the DFT orbital energies. The matrix elements describe the coupling of single-particle transitions from occupied orbitals $i,j$ to unoccupied orbitals $a,b$, mediated by the two-electron Coulomb interaction $(ia|jb)$.
The transition moments are computed from the eigenvectors $X^{\nu}/Y^{\nu}$ of Eq.~\eqref{eq:rpa_eigenproblem} as
\begin{equation}
  \rho^\nu_{cc} = \sum_{ia} (cc|ia)(X^\nu_{ia} + Y^\nu_{ia})
  \label{eq:transition_amplitudes}
\end{equation}
and contain the Coulomb coupling between an ionized level $c$ and a charge-neutral excitation $\nu$. The  energies $\epsilon_c^{\text{QP}}$ are computed from the $GW$ self-energy, which is constructed from the eigenvalues and eigenvectors of Eq.~\eqref{eq:rpa_eigenproblem}.
\par
The poles of $G^\mathrm{QP}_{c}$ (Eq.~\eqref{eq:qp_propagator}) and $G^1_{c}$ (Eq.~\eqref{eq:first_order_satellite}) determine the positions of QP and satellite peaks in the spectral function, located at $\omega_c^{\text{QP}}=\epsilon_c^{\text{QP}}$ and $\omega^{\text{Sat},\nu}_c=\epsilon^\mathrm{QP}_c-\Omega^{\nu}$, respectively. In $GW+C$, the energy separation between each satellite and the QP peak equals thus exactly the excitation energies $\Omega^\nu$. Since the theoretical framework is formulated on a negative energy scale, the binding energies $\varepsilon_c$ are obtained by multiplying by $-1$, e.g., $\varepsilon^\mathrm{QP}_c = -\epsilon^\mathrm{QP}_c$. 

\subsection{Relation between optical and shake-up spectra}
\label{sec:transition_densities}
In XPS, charge-neutral valence excitations manifest as satellite features that emerge alongside the ionization process. The same type of excitations can also be probed directly by UV–vis absorption spectroscopy. In our theoretical framework (at its current stage of development), the energies of shake-up and optical excitations coincide, which is a reasonable approximation for excitations with predominantly $\pi \rightarrow \pi^*$ character.\cite{brisk1975shake,kocklauner2025gw, rocco2008electronic, zhang2024core} Based on this premise, shifting the XP spectrum - so that the main ionization peak is set to zero energy - aligns the satellite positions with the optical excitation energies. Their intensities, however, differ strongly because the underlying mechanisms are governed by different selection rules.
\par
Optical excitations couple directly to the electromagnetic field of the incident light. Their intensity is quantified by the oscillator strength $f^\nu$, typically evaluated within the dipole approximation,\cite{casida2012progress,norman2018simulating} 
\begin{equation}
  f^\nu = \frac{2}{3} \Omega^\nu \left|\int \text{d}\mathbf{r} \, \hat{\mathbf{r}}\,n_\nu(\mathbf{r})\right|^2
  \label{eq:oscillator_strength}
\end{equation}
where the integral represents the transition dipole moment expressed through the position operator $\hat{\mathbf{r}}$ and the transition density $n_\nu(\mathbf{r})$. The latter is given by\cite{jornet2019real}
\begin{equation}
  n_\nu(\mathbf{r}) = \sum_{ia} \left(X^\nu_{ia} + Y^\nu_{ia}\right) \psi_i(\mathbf{r}) \psi_a(\mathbf{r})
  \label{eq:transition_density}
\end{equation}
where $\psi_i$ and $\psi_a$ are initial (occupied) and final state (unoccupied) orbitals and $X^\nu_{ia}$, $Y^\nu_{ia}$ excitation amplitudes. If we stay in the RPA framework, $\Omega^{\nu}$ and $X^\nu/Y^\nu$ are obtained by solving  Eq.~\eqref{eq:rpa_eigenproblem}.
\par
Shake-up excitations do not couple to the electromagnetic field of the incident light but are triggered by the ionization process.\cite{martin1976theory} Within $GW+C$, their intensities $\gamma_c^{\nu}$ (Eq.~\eqref{eq:gamma_definition}) are determined by the transition moments $\rho^\nu_{cc}$ (Eq.~\eqref{eq:transition_amplitudes}), which can be rewritten in terms of the core-level charge density $n_{c}(\mathbf{r}) = |\psi_{c}(\mathbf{r})|^2$ and the transition density $n_\nu(\mathbf{r})$:
\begin{equation}
  \rho^\nu_{cc} = \int \text{d}\mathbf{r} \, \text{d}\mathbf{r}^\prime \, \frac{n_{c}(\mathbf{r}) \, n_\nu(\mathbf{r}^\prime)}{|\mathbf{r} - \mathbf{r}^\prime|}
\label{eq:core_transition_amplitude}
\end{equation}
This form shows that charge-neutral excitations couple to the core hole (a single charge, or monopole) through the Coulomb operator $|\mathbf{r}-\mathbf{r}'|^{-1}$, unlike optical excitations, whose intensities are determined by the dipole operator. To highlight this distinction, we refer to the rules implied by Eq.~\eqref{eq:oscillator_strength} as dipole selection rules, and those implied by Eq.~\eqref{eq:core_transition_amplitude} as monopole selection rules.
\par
The selection rules determine whether the transition amplitudes $f^\nu$ and $\rho_{cc}^\nu$ vanish or remain finite and are derived from the symmetry and spatial properties of the operators and integrated quantities.
The position operator in Eq.~\eqref{eq:oscillator_strength} has odd symmetry, and the transition density is integrated across the entire molecule. In contrast, the Coulomb operator in Eq.~\eqref{eq:core_transition_amplitude} has even symmetry,
and the transition moments are highly sensitive to spatial overlap between the localized $n_c(\mathbf{r})$ and the excitation's transition density.
Consequently, excitations that are bright in optical spectra may be weak or dark in satellite spectra, and vice versa as shown in Section~\ref{sec:h2tpp}.

\section{Systems and Methods}
\label{sec:systems_and_methods}
\subsection{Molecular Systems}
\label{sec:molecules}
\begin{figure*}[!h]
    \centering
    \includegraphics[width=0.95\linewidth]{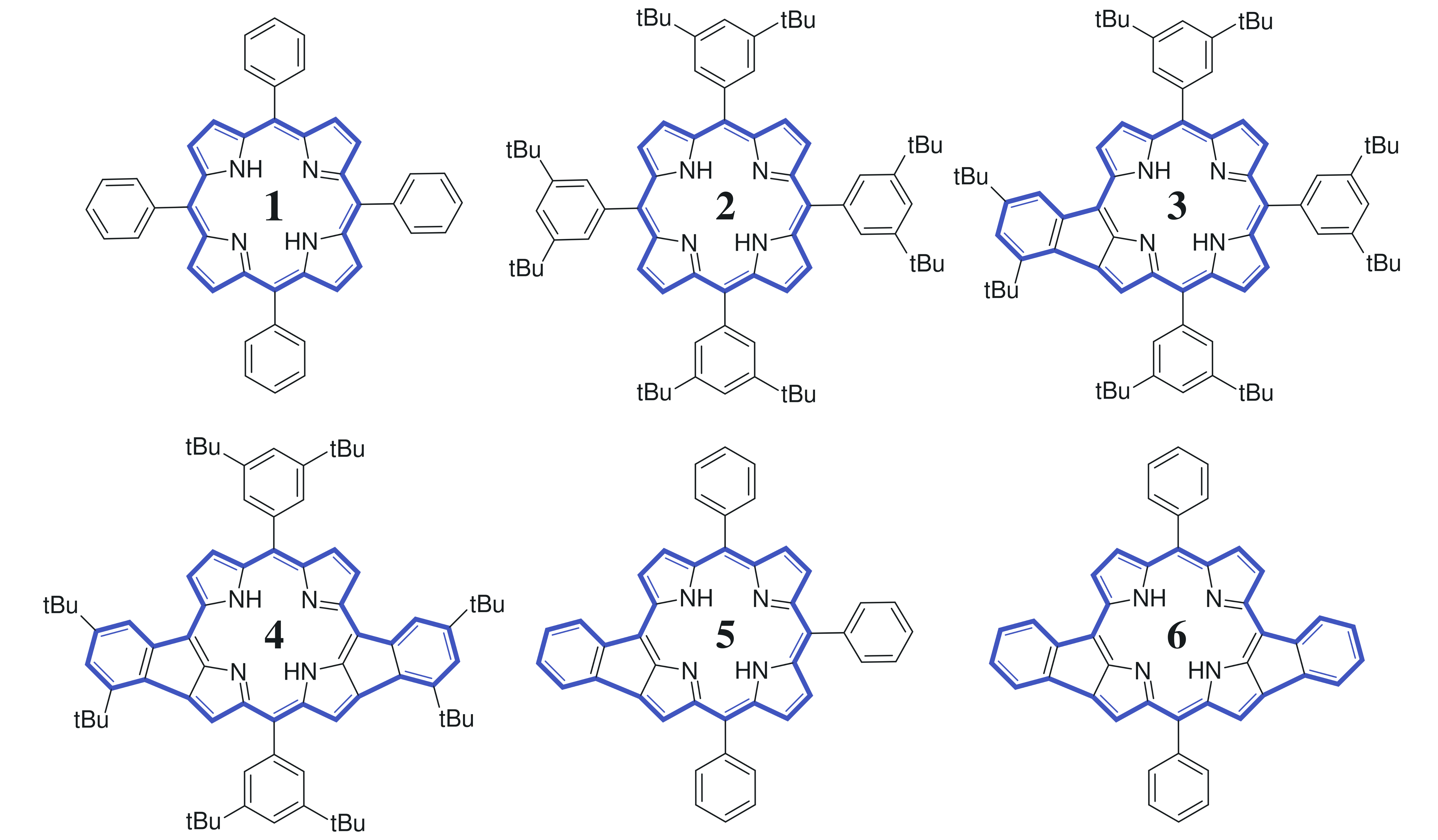}
    \caption{Summary of the porphyrin molecules considered in this study. The conjugated macrocycle is highlighted in blue. Systems \textbf{1-4} have been realized experimentally (see Refs.~\citenum{oleszak2024fused} and \citenum{martin2020oxidative}), while \textbf{5} and \textbf{6} are experimentally not realizable, but serve as smaller model systems for the computational analysis of compounds \textbf{3} and \textbf{4}.}
    \label{fig:molecules}
\end{figure*}
In this work, we study the series of tetraphenylporphyrin-based molecules displayed in Figure~\ref{fig:molecules}.
The smallest molecule is free-base tetraphenylporphyrin (H$_2$-TPP, \textbf{1}). Molecule \textbf{2} is a variant of H$_2$-TPP with additional tert-butyl side groups. Molecules \textbf{3} and \textbf{4} were only recently synthesized~ \cite{martin2020oxidative} and are the mono- and doubly fused derivatives of \textbf{2}, in which one (\textbf{3}) or two (\textbf{4}) phenyl rings are fused to the macrocycle. Molecules \textbf{5} and \textbf{6} are simplified analogues of the fused derivatives, obtained by removing the tert-butyl groups. Compounds \textbf{1}–\textbf{4} were experimentally synthesized and measured with XPS, while all six compounds were computationally studied.
\par
The choice of these systems is determined by practical limitations in the synthesis of these molecules: Direct ring fusion of the parent compound \textbf{1} is not possible, necessitating the introduction of bulky substituents such as tert-butyl groups in \textbf{2} to exploit steric effects. While these groups facilitate ring fusion, they also significantly increase the molecular size by roughly 100 atoms going  from \textbf{1} to \textbf{2}.
\par
While we can compute the XP spectra of tert-butyl porphyrins \textbf{2}-\textbf{4}, these molecules are too large for a detailed analysis of individual excitations, which is explained in more detail in Section~\ref{sec:comp_details}. To overcome this limitation, we introduce the smaller series \textbf{1}, \textbf{5}, and \textbf{6}, which likewise progress from unfused to mono-fused to doubly fused.  Although \textbf{5} and \textbf{6} are only computational models and not experimentally realizable, they allow us to probe the qualitative electronic effects of ring fusion without the computational cost imposed by bulky substituents.

\subsection{Computational Details}
\label{sec:comp_details}

Implementing the $GW+C$ as outlined in Section~\ref{sec:gw_theory}, leads to an $O(N^6)$ scaling with respect to system size $N$, originating from the diagonalization of the Casida matrix in Eq.~\eqref{eq:rpa_eigenproblem}. This limits the tractable system size to $<100$ atoms. To overcome this bottleneck, we recently developed a $GW+C$ scheme with $O(N^4)$ scaling~\cite{kocklauner2025gw} and implemented it in the all-electron FHI-aims program package,\cite{blum2009ab, abbott2025roadmap} which is based on numeric atom-centered orbitals. Together with our latest performance optimizations, our implementation enables calculations on molecules of up to 200 atoms.

For our low-scaling $GW+C$ implementation we rewrite $C_c(t)$ in Eq.~\eqref{eq:cumulant_propagator} by the following integral over the correlation part of the $GW$ self-energy $\Sigma_c(\omega)$
    \begin{equation}
    \label{eq:numerical_cumulant}
    C_c(t) = \frac{1}{\pi}\int \mathrm{d}\omega \, \frac{\left| \mathrm{Im} \Sigma_c(\omega + \epsilon_c) \right|}{\omega^2} \left(e^{-\mathrm{i}\omega t} + \mathrm{i}\omega t - 1\right) 
\end{equation}
where $\epsilon_c$ is the DFT orbital energy of the core level. 
This integral is solved numerically using a dense frequency grid. We obtain the $GW$ self-energy numerically, using the contour-deformation (CD) in combination with the analytical continuation of $W$ (CD-WAC)~\cite{panades2023accelerating,Leucke2025}. 
The final spectral function is computed by taking the imaginary part of the Fourier transform of Eq.~\eqref{eq:cumulant_propagator} following Eq.~\eqref{eq:spectral_function}. \par

Our low-scaling $GW+C$ scheme provides the correct positions and intensities for both the main excitations and the satellites. It also allows satellites to be assigned to individual core-levels as demonstrated in Ref.~\citenum{kocklauner2025gw}. However, the dissection of the satellite spectrum in individual transitions $\nu$ or the analysis of the transition density $n_{\nu}(\mathbf{r})$ requires the knowledge of the eigenvectors $X^\nu, Y^\nu$ and eigenvalues $\Omega^\nu$, which are only available through the computationally expensive fully analytic scheme (Eqs.~\eqref{eq:gamma_definition} to \eqref{eq:transition_amplitudes}), where “fully analytic” means that only matrix algebra is required and no numerical frequency integrals are evaluated.\par

For the purpose of deeper analysis, we also implemented the $O(N^6)$-scaling analytic $GW+C$ scheme as part of this work into the FHI-aims package, using the existing linear response TDDFT implementation~\cite{liu2020all} as a starting point. In addition, we implemented the $GW$ scheme~\cite{golze_gw_2019} in analytic form to obtain the QP energies $\epsilon_c^{\text{QP}}$ (Eqs.~\eqref{eq:qp_propagator} and \eqref{eq:first_order_satellite}). In this approach, the $GW$ self-energy is constructed directly from the eigenvectors $X^\nu, Y^\nu$ and eigenvalues $\Omega^\nu$, thereby avoiding numerical approximations such as the CD. Using MPI parallelization, our fully analytic implementation is applicable to systems with up to 80 atoms. 
These calculations were performed for compounds \textbf{1}, \textbf{5} and \textbf{6} (max. 78 atoms).
\par
The broadening of the $GW+C$ spectral function, controlled by the parameter $\eta$, is purely Lorentzian, as it contains lifetime effects, but no extrinsic and instrumental broadening effects which introduce additional Gaussian broadening.
To compare $GW+C$ spectral functions with the experiment, the theoretical spectra were calculated with a very narrow intrinsic broadening of $\eta=0.1$~eV and, in a post-processing step, further broadened to fit the experimental data using a 80:20 Gaussian-Lorentzian filter with a width of $0.42$~eV.
\par
For the $GW+C$ calculations, we used the settings identified as optimal in Ref.\citenum{kocklauner2025gw}. A detailed summary of the calculation parameters is provided Section~S1 in the Supporting Information (SI). The geometries as well as all input and output files for the FHI-aims calculations are available in the NOMAD database.\cite{nomad_reference}

\subsection{Experimental Details}
The XPS measurements were performed for thick multilayers of the different porphyrins with an SES 200 hemispherical analyzer and a SAX 100 monochromatized Al K$\alpha$ X-ray source, producing a Au 4f$_{7/2}$ line with a full-width-at-half-maximum (FWHM) of 1.04~eV.  The porphyrin multilayers (thickness $>$10~ML) were deposited using an organic evaporator onto a copper substrate held at room temperature. In order to achieve good statistics, the N~1s level was measured with a total measuring time of 14-16~h while taking great care to avoid beam damage. 

\section{Results and Discussion} 
\label{sec:results}
We start by analyzing how the extension of the $\pi$-electron system through ring fusion is reflected in the shake-up satellite spectrum by comparing experimental XP spectra of Porphyrin multilayers with theoretical calculations and using the latter to assign the satellites to individual core ionizations. We then provide a detailed theoretical analysis of the charge-neutral valence excitations responsible for the satellites, explaining the observed changes upon ring fusion. 

\subsection{Sensitivity of satellite features to structural modifications}
\label{sec:results_experiment}
\begin{figure*}[!h]
    \centering
    \includegraphics[width=\linewidth]{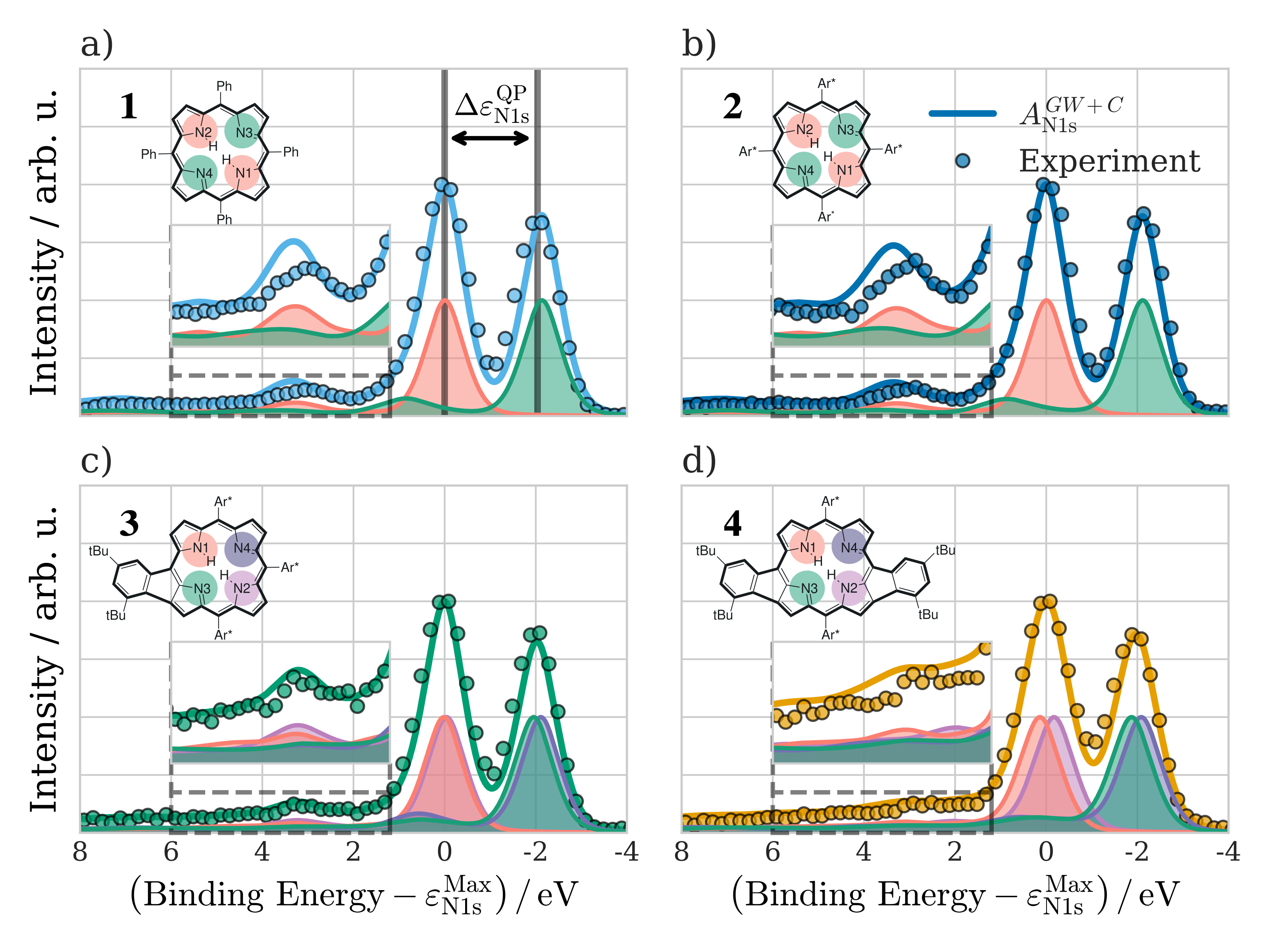}
    \caption{Comparison of the experimental XP spectrum (dots) alongside the computed spectral function $A_\mathrm{N1s}^{GW+C}(\omega)$ (solid line) for molecules \textbf{1-4} (a-d), with the inset zooming into the region of the pyrrolic satellite around $3.0-3.5$~eV by a factor of 8. All spectra are aligned at the energy of maximum intensity in the N~1s spectrum $\varepsilon_\mathrm{N1s}^\mathrm{Max}$. For $A_\mathrm{N1s}^{GW+C}(\omega)$, the individual spectra of the pyrrolic (pink, violet) and iminic (green, purple) N~1s levels are indicated by the shaded areas. The respective nitrogen atoms are highlighted in the structures, with functional groups abbreviated as Ph=Phenyl and Ar\textsc{*}=3,5-di-\textit{tert}-butyl-phenyl. }
    \label{fig:experiment_comparison}
\end{figure*}

\begin{table*}[!h]
    \fontsize{10}{12}\selectfont
    \centering
    \begin{tabular}{c|cccc|cccccccc}
    \toprule
       \multicolumn{1}{c}{\textbf{Molecule}}  & \multicolumn{4}{c}{\textbf{Experiment}} & \multicolumn{7}{c}{\textbf{Theory}} \\
       \midrule
        & $\Delta \varepsilon^\mathrm{QP}_\mathrm{N1s}$& $I_1 / I_2$& $ \varepsilon^\mathrm{Max}_\mathrm{N1s}$&$\Delta \varepsilon_\mathrm{N1/N2}^\mathrm{Sat}$&$\Delta \varepsilon^\mathrm{QP}_\mathrm{N1s}$& $I_1 / I_2$& $\varepsilon^\mathrm{Max}_\mathrm{N1s}$& $\Delta \varepsilon_\mathrm{N1}^\mathrm{Sat}$ &  $\Delta \varepsilon_\mathrm{N2}^\mathrm{Sat}$&  $\Delta \varepsilon_\mathrm{N3}^\mathrm{Sat}$&  $\Delta \varepsilon_\mathrm{N4}^\mathrm{Sat}$\\
       \midrule
        \textbf{1} &  2.06  & 1.26 &400.14 & 3.10 & 2.13 &    1.23 &  404.69&  3.28 & 3.28 & 2.98 & 2.98  \\
        \textbf{2}&  2.11 & 1.22  &400.12& 2.98 &2.12 &1.23 & 404.43& 3.30 & 3.30 & 2.98& 2.98 \\
        \textbf{3}&  2.01 & 1.24 &400.11& 3.00 &2.03 & 1.25 & 404.41& 3.25 &3.27 &2.46 & 2.69  \\
        \textbf{4}& 1.96 & 1.23 &399.88& - &1.97 & 1.24 &  404.40& 2.95& 2.15 & 2.09 & 2.64\\
         \bottomrule
    \end{tabular}
    \caption{Key spectral parameters extracted from experimental and theoretical N~1s XP spectra, including the energy separation $\Delta \varepsilon^\mathrm{QP}_\mathrm{N1s}$ and intensity difference $I_1/I_2$ between the two main peaks as well as the binding energy $\varepsilon^\mathrm{Max}_\mathrm{N1s}$ of the peak with maximum intensity. We also report the separation between the photoionization peak and the most prominent satellite of each N atom, $\Delta\varepsilon_{\mathrm{N}X}^\mathrm{Sat} = \varepsilon_{\mathrm{N}X}^\mathrm{Sat} - \varepsilon_{\mathrm{N}X}^\mathrm{QP}$, where $\varepsilon_{\mathrm{N}X}^\mathrm{Sat}$ denotes the satellite binding energy. All energy values in eV.}
\label{tab:spectrum_properties}
\end{table*}

In Figure~\ref{fig:experiment_comparison}, we compare the experimental N~1s XP spectra of compounds \textbf{1}-\textbf{4} with the $GW+C$ spectral functions $A^{GW+C}_\mathrm{N1s}$. To facilitate comparison, all spectra are aligned to the binding energy with maximum intensity, $\varepsilon_\mathrm{N1s}^\mathrm{Max}$.
Each spectrum includes the contributions from all four nitrogen atoms. Theoretical predictions of individual contributions are shown as semi-transparent overlays, with the color of each curve matching the corresponding atom highlighted in the molecular structure inset of Figure~\ref{fig:experiment_comparison}.
\par
The spectra of compounds \textbf{1}-\textbf{4} share a similar overall shape, featuring two main peaks separated by $\Delta \varepsilon^\mathrm{QP}_\mathrm{N1s}$, as indicated in Figure~\ref{fig:experiment_comparison}a. The peak at higher binding energy corresponds to the protonated pyrrolic (N1, N2) and the peak at lower binding energy to the unprotonated iminic (N3, N4) nitrogen atoms. For compounds \textbf{1}-\textbf{3} we observe two satellites.
One satellite is located at an energy separation of approximately 3 eV toward higher binding energy relative to the pyrrolic peak.
The second satellite overlaps with the pyrrolic main peak, increasing its intensity ($I_1$) relative to the iminic peak ($I_2$), as reflected in the intensity ratio $I_1/I_2\approx1.2$.\cite{macquet1978x, gottfried2015surface}
\par
Comparing the spectra of \textbf{1}-\textbf{4},
we find that the main lines are largely unaffected by the structural modifications. As summarized in Table~\ref{tab:spectrum_properties}, the relative splitting and absolute binding energies are essentially the same: $\Delta \varepsilon^\mathrm{QP}_\mathrm{N1s}$ deviates by less than 0.1~eV, and $\varepsilon_\mathrm{N1s}^\mathrm{Max}$ deviates by no more than 0.25~eV. The intensity ratios $I_1/I_2$ remain virtually unchanged. Consequently, distinguishing compounds \textbf{1}–\textbf{4} in XPS based on the photoionization peaks is not possible.
\par
Comparison of the satellite features for \textbf{1} and \textbf{2} shows that both exhibit a distinct satellite approximately 3~eV from the main pyrrole peak; thus, replacing the phenyl group in \textbf{1} with 3,5-di-\textit{tert}-butylphenyl has no impact on these features. However, unlike the main lines, the satellite spectrum is clearly affected by ring fusion: The satellite at $\varepsilon_{\mathrm{N1s}}^{\mathrm{Max}}+ 3$~eV becomes less pronounced for \textbf{3} and disappears in \textbf{4}, where it is replaced by a broad tailing feature. The higher sensitivity of the shake-up satellite excitations to structural modifications is due to the fact that, in addition to core ionization, we also probe the delocalized frontier orbitals, both occupied and unoccupied. Consequently, any changes in these orbitals will be reflected in the satellite features. 
\par
We observe excellent agreement between experimental data and the the $GW+C$ spectral function across all four molecules:
The splitting $\Delta \varepsilon^\mathrm{QP}_\mathrm{N1s}$ is predicted within 0.1~eV of the experimental value (see Table\ref{tab:spectrum_properties}). The intensity ratios $I_1/I_2$ are reproduced equally well, differing by at most 3~\% from experiment. The $GW+C$ theory correctly predicts that the main lines for \textbf{1}–\textbf{4} occur at approximately the same energy. However, the absolute energies $\varepsilon_\mathrm{N1s}^\mathrm{Max}$ differ by $\approx 4$~eV from experiment because the calculations were performed in the gas phase, whereas the experimental spectra are solid-state XPS, where the reference is not the vacuum level but the Fermi level of the sample and thus the spectra are shifted by the work function.
\par
For the unfused porphyrins \textbf{1} and \textbf{2} (Figure~\ref{fig:experiment_comparison}a,b), the position and intensity of the satellite at $\varepsilon_{\text{N1s}}^{\text{Max}}+3.3$ eV agree well with experiment, with deviations of about 0.2–0.3 eV. 
In the partly fused systems \textbf{3} (Figure~\ref{fig:experiment_comparison}c), we correctly predict the reduced satellite intensity compared to \textbf{1} and \textbf{2}. As in experiment, this satellite nearly vanishes in the XP spectrum of \textbf{4} (Figure~\ref{fig:experiment_comparison}d), giving way to a broad tailing feature.
\par
For the theoretical spectra, deconvolution into the individual core-level contributions is straightforward.
For compounds \textbf{1} and \textbf{2}, the nitrogen atoms form the two chemically equivalent pairs N1/N2 and N3/N4. 
The atoms in each pair yield not only indistinguishable main features but also indistinguishable satellite features.
The iminic pair N3/N4 gives rise to the satellite that overlaps with the pyrrolic photoionization peak, whereas the pyrrolic pair N1/N2 produces the distinct satellite feature appearing approximately 3~eV from the corresponding main line.
\par
For the fused compounds \textbf{3} and \textbf{4}, the symmetry is broken, which affects the spectral features in two ways. First, the previously equivalent main lines split slightly. This splitting is negligible in the spectral function of \textbf{3} and is about 0.3~eV for \textbf{4}, which is not resolved in the experiment. Second, and more significantly, the symmetry breaking gives rise to four distinct satellite spectra, i.e., one for each nitrogen atom.
\par
For compound \textbf{3}, the pyrrolic satellites have both approximately the same binding energy (see also $\Delta\varepsilon_{\mathrm{N}1}^\mathrm{Sat}$ and $\Delta\varepsilon_{\mathrm{N}2}^\mathrm{Sat}$ in Table~\ref{tab:spectrum_properties}), although the N1 satellite is less intense than N2 satellite (see inset of Figure~\ref{fig:experiment_comparison}c). Compared to \textbf{1} and \textbf{2}, the total spectrum still exhibits a distinct, but smaller pyrrolic satellite. 
However, in compound \textbf{4}, the differences between the individual satellite spectra become pronounced.
In particular, the maxima of the pyrrolic satellites differ by as much as 0.8~eV (see $\Delta\varepsilon_{\mathrm{N}1/\mathrm{N}2}^\mathrm{Sat}$, Table~\ref{tab:spectrum_properties}), which adds up with the symmetry breaking of the main lines to a total pyrrolic satellite splitting of 1.1~eV and results in a broad, unresolved shoulder in the total spectrum instead of a distinct satellite peak.
\par
The different splitting of the main and satellite peaks (0.3~eV vs 1.1~eV) after symmetry breaking by fusion indicates that the pyrrolic core levels N1 and N2 no longer couple to the same charge-neutral excitations but to different ones, which will be analyzed in detail in the next section.

\subsection{Tracing the origin of shake-up satellites}
\label{sec:results_transition}
To understand \textit{why} the satellite spectrum changes upon ring fusion, we analyze the charge-neutral valence excitations that strongly couple to core-level ionization. 
In Section~\ref{sec:h2tpp}, we present a detailed analysis for \textbf{1} (H2-TPP), which serves as a representative of the unfused porphyrins. As described in Section~\ref{sec:molecules}, we use the smaller analogue \textbf{5} as a model for the partially fused porphyrin \textbf{3}, and \textbf{6} as a model for the fully fused compound \textbf{4}. The analysis of \textbf{6} is given in Section~\ref{sec:fused}, while that of \textbf{5} is provided in Section~S2 of the SI.
Although the satellite spectra of the smaller analogues show reduced quantitative agreement with experiment, the qualitative comparison with their parent compounds remains meaningful (see Figure~S3 in the SI).
\subsubsection{Unfused porphyrin H2-TPP}
\label{sec:h2tpp}
\begin{figure*}[!h]
    \centering
    \includegraphics[width=\linewidth]{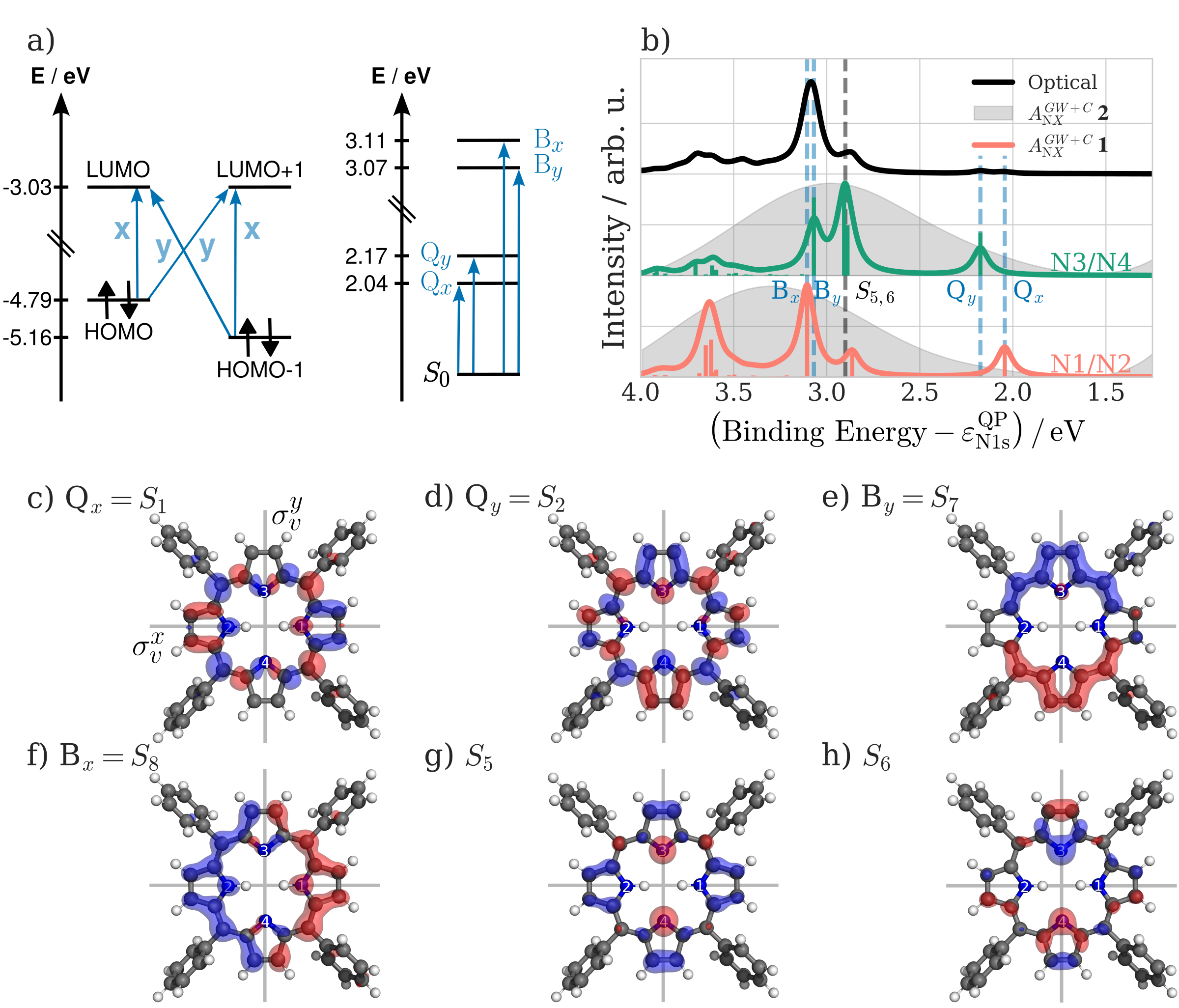}
    \caption{a) Summary of the four-orbital Gouterman model for valence excitations in Porphyrins. b) Direct comparison of the first-order N~1s satellite spectra (colored  lines) of the symmetry-equivalent N1/N2 and N3/N4 states with the optical absorption spectrum (black line). Individual satellite intensities $\gamma^\nu_\mathrm{N1s}$ are shown as stick spectrum. Excitations with large intensities in the satellite spectrum are highlighted by the dotted lines and labeled following the Gouterman model if possible. The maximum of each spectrum is set to the  same value to enable  comparison. The respective broadened $GW+C$ satellite spectra of \textbf{2} (from Figure~\ref{fig:experiment_comparison}) are shown in gray as reference. c-h) show iso-surfaces of transition densities $n_\nu(\mathbf{r})$ for all excitations highlighted in b). The two possible mirror planes $\sigma_v^{x,y}$, which are perpendicular to the molecular plane, are shown in grey.}
    \label{fig:h2tpp_transition_densitites}
\end{figure*}

\begin{table}[!h]
    \fontsize{10}{12}\selectfont
    \centering
    \begin{tabular}{ccccc}
    \toprule \textbf{State}  & $\Omega^\nu$ / eV & $f^\nu $& $\gamma^\nu_\mathrm{N1/N2}$&$\gamma^\nu_\mathrm{N3/N4}$  \\
       \midrule
        $S_1$ (Q$_x$) & 2.04& 0.028& 0.020 &  0.000\\
        $S_2$ (Q$_y$) & 2.17& 0.043& 0.000 & 0.027\\
        $S_4$ & 2.86 & 0.212 & 0.015 & 0.000 \\
        $S_5$ & 2.89& 0.003& 0.000 &  0.032\\
        $S_6$ & 2.91& 0.086& 0.000 & 0.059 \\
        $S_7$ (B$_y$) & 3.07& 0.907& 0.000 & 0.050\\
        $S_8$ (B$_x$) & 3.11& 0.734& 0.060 & 0.000\\    
        \bottomrule
    \end{tabular}
    \caption{Theoretical RPA excitation energies $\Omega^\nu$, dimensionless oscillator strength $f_\nu$ (Eq.~\eqref{eq:oscillator_strength}) compared to the dimensionless satellite intensities $\gamma^\nu_\mathrm{N1s}$ (Eq.~\eqref{eq:gamma_definition}) in H2-TPP (\textbf{1}). Satellite intensities are averaged across the approximately symmetry-equivalent pyrrolic ($\gamma^\nu_\mathrm{N1/N2}$) and iminic ($\gamma^\nu_\mathrm{N3/N4}$) core-levels.}
    \label{tab:h2tpp_intensities}
\end{table}

For porphyrins like \textbf{1}, valence excitations are commonly described by the Gouterman model\cite{gouterman1963spectra} summarized in Figure~\ref{fig:h2tpp_transition_densitites}a, which explains their characteristic optical absorption spectra. The model involves only four frontier orbitals— the two highest occupied molecular orbitals (HOMO and HOMO-1) and the two nearly degenerate lowest unoccupied molecular orbitals (LUMO and LUMO + 1). As shown in Figure~\ref{fig:h2tpp_transition_densitites}a, transitions between these orbitals can couple in two distinct ways labeled as $x$ and $y$ components. $x$ excitations are linear combinations of HOMO$\rightarrow$ LUMO and HOMO-1$\rightarrow$ LUMO+1 singlet transitions, whereas $y$ excitations combine HOMO$\rightarrow$ LUMO+1 and HOMO-1$\rightarrow$ LUMO transitions. Different linear combinations are possible: Constructive interference between the transition dipole moments yields the optically  bright higher-energy B$_x$ and B$_y$ (Soret) bands, whereas destructive interference yields less intense lower-energy Q$_x$ and Q$_y$ states.
\par
In Figure~\ref{fig:h2tpp_transition_densitites}b, we compare the theoretical optical absorption spectrum with the first-order satellite spectra ($G_\mathrm{N1s}^1$, see Eq.~\eqref{eq:first_order_satellite}) of  the pyrrolic (N1/N2) and iminic (N3/N4) core levels. For convenience, we add the broadened theoretical spectrum of the larger compound \textbf{2} from Figure~\ref{fig:experiment_comparison}a in gray. Both the optical and satellite spectra are obtained at the RPA level of theory. As shown in our previous work~\cite{kocklauner2025gw}, RPA yields reasonably good charge-neutral excitation energies for low-lying excitations of $\pi\rightarrow\pi^*$ character, despite the absence of exchange and correlation coupling terms in Eq.~\eqref{eq:rpa_matrices}. We also find that RPA is sufficiently accurate for the porphyrins, as demonstrated by the good agreement between the theoretical and experimental~\cite{oleszak2024fused} absorption spectra and with more accurate theoretical data based on the Bethe-Salpeter equation (BSE), see Figure~S4 in the SI.\par
For the optical spectrum, ab initio calculations with RPA yield numerous singlet excitations, rather than the two Q- and two B-bands predicted by the simplistic Gouterman model. The eight lowest-energy excitations ($S_1 - S_8$) are reported in Table~\ref{tab:h2tpp_intensities}, except for $S_3$, which is omitted due to zero intensity. The most intense peak corresponds to the Soret bands B$_x$ and B$_y$ around 3~eV, while Q-bands  at $\approx 2$~eV are almost dark. However, an additional excitation with considerable spectral weight is observed at 2.8~eV ($S_4$, see Table~\ref{tab:h2tpp_intensities}), which is not captured by the Gouterman model. 
\par
Although the excitation energies of the $GW+C$ satellites are the same as in the optical spectrum, the intensity of the satellite bands differs strongly from their optical intensities. Notably, the most intense feature in the N3/N4 spectrum originates from the optically weak or almost dark $S_5$ and $S_6$ transitions at approximately 2.9~eV (see Table~\ref{tab:h2tpp_intensities}), while the optically bright $S_4$ state has no contribution.
Furthermore, we find that the satellite spectra of the N1/N2 and N3/N4 nitrogens are very different from each other, with mutual exclusion rules apparently in effect for the Gouterman excitations:
The N1/N2 spectrum contains only satellites stemming from the Q$_x$ and B$_x$ excitations, while the N3/N4 satellite spectrum features the Q$_y$ and B$_y$ bands. Furthermore, $S_5$ and $S_6$ appear in the N3/N4 spectrum, but not in N1/N2.  
\par
The key quantity to understand the intensity differences between satellite and optical spectra, as well as the differences between the iminic and pyrrolic satellites, is the transition density $n_\nu(\mathbf{r})$ (Eq.~\eqref{eq:transition_density}). This quantity measures the overlap of initial and final states for a given excitation $\nu$. Qualitatively, the intensity $\gamma^{\nu}$ of a satellite (Eqs.~\eqref{eq:gamma_definition}) is governed by two main criteria: (1) the symmetry of $n_{\nu}(\mathbf{r})$ must satisfy the monopole selection rule and (2) $n_{\nu}(\mathbf{r})$ must be significant in the vicinity of the core-hole.
In Figure~\ref{fig:h2tpp_transition_densitites}c-h, we show the transition densities for the six most intense low-energy shake-up satellites, also indicated by vertical dashed lines in Figure~\ref{fig:h2tpp_transition_densitites}b. All excitations are \(\pi \rightarrow \pi^*\) transitions delocalized across the conjugated \(\pi\)-system. Blue iso-surfaces indicate spatial regions where $n_{\nu}$ is negative and red where it has a positive sign.
\par
First, we will analyze the satellites in terms of criterion 1, the symmetry criterion:
A transition is symmetry-forbidden if the product of the N~1s core-level electron density and the transition density in Eq.~\eqref{eq:core_transition_amplitude}, $n_\mathrm{N1s}(\mathbf{r}) n_\nu(\mathbf{r})$, is antisymmetric with respect to a shared symmetry element. In that case, the integral in Eq.~\eqref{eq:core_transition_amplitude} vanishes, or in other words, the monopole selection rule is not fulfilled. 
The symmetry of the product $n_\mathrm{N1s}(\mathbf{r})n_\nu(\mathbf{r})$ can be most effectively analyzed in terms of the relevant symmetry groups and elements.\par

The (approximate) symmetry group of the macrocycle is $D_{2h}$ for the ground state, which is lowered to $C_{2v}$ upon creation of a localized core-hole at one distinct nitrogen site. In $C_{2v}$, one mirror plane lies in the macrocycle plane and the other is perpendicular to it. For a core hole at N1 or N2, the perpendicular plane is $\sigma_v^x$, while for N3 or N4 it is $\sigma_v^y$, see Figure~\ref{fig:h2tpp_transition_densitites}c.
The mirror planes $\sigma_v^x$ and $\sigma_v^y$ contain the N1/N2 and N3/N4 atoms, respectively.  
The 1s electron density $n_\mathrm{N1s}(\mathbf{r})$ is strictly positive (i.e. phase-less) and localized at the respective N atom and thus always symmetric with respect to these mirror planes. Consequently, it is the symmetry of the transition density that determines whether an excitation is dark or bright.

\par
As shown in Figure~\ref{fig:h2tpp_transition_densitites}c-h, the transition densities of Q\(_y\), B\(_y\) and S$_6$ are antisymmetric with respect to $\sigma_v^x$. Consequently, these excitations have zero intensity in the satellite spectra of N1/N2, see also Table~\ref{tab:h2tpp_intensities}. Similarly, the transition densities of Q\(_x\) and B\(_x\) are antisymmetric with respect to $\sigma_v^y$ and therefore these excitations do not appear for N3/N4. The mutual exclusion rule observed for the Q- and B-bands is a direct consequence of the fact that the Q$_x$/Q$_y$ and B$_x$/B$_y$ transition densities are related by symmetry. For example, the B$_x$ transition density can, to a good approximation, be mapped onto B$_y$ by a $\pi/2$ rotation around the intersection axis of $\sigma_v^x$ and $\sigma_v^y$, with an analogous correspondence between Q$_x$ and Q$_y$.\par

\begin{figure*}[ht!]
    \centering
    \includegraphics[width=\linewidth]{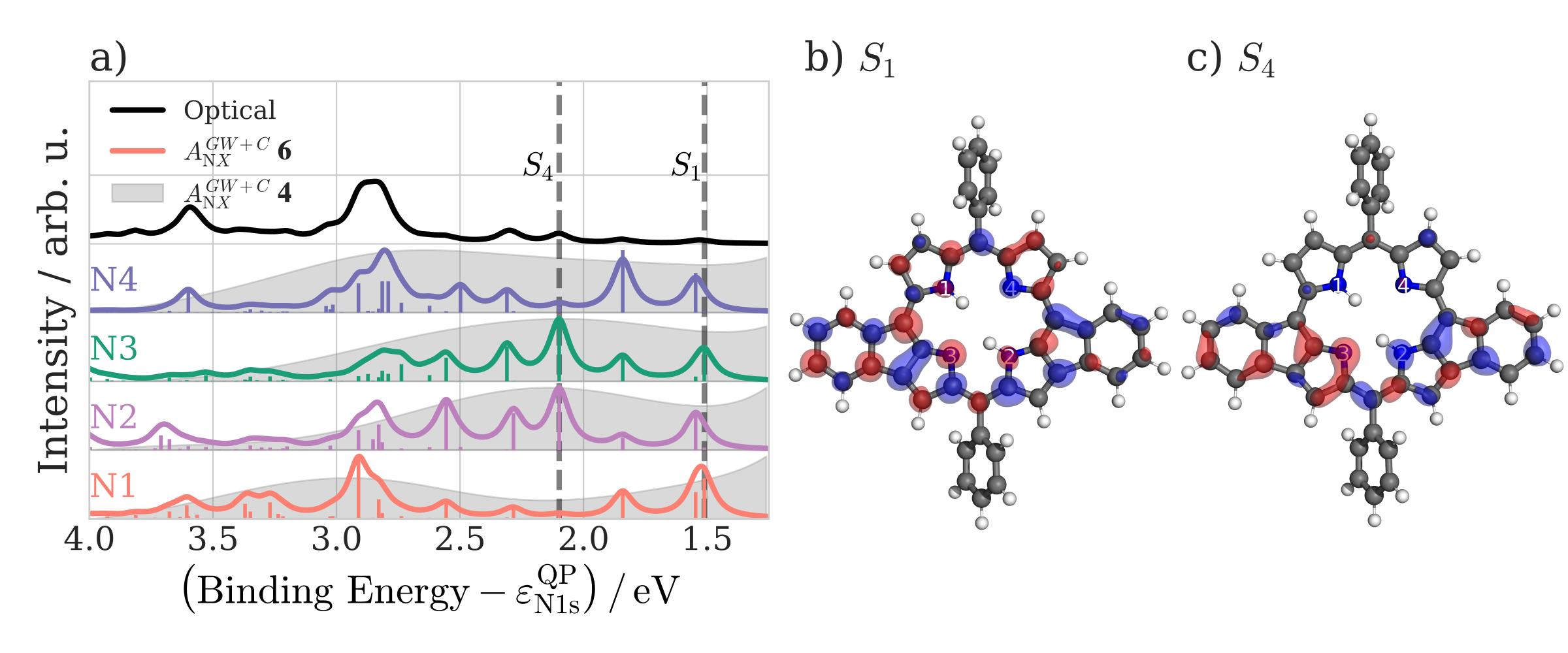}
    \caption{a) Direct comparison of the first-order N~1s satellite spectra (colored  lines) of all N~1s states with the optical absorption spectra (black line). Individual satellite intensities $\gamma^\nu_\mathrm{N1s}$ are shown as stick spectrum. The maximum of each spectrum is set to the same value to enable comparison. The respective broadened $GW+C$ satellite spectra of \textbf{4} (from Figure~\ref{fig:experiment_comparison}) are shown in gray as reference. b,c) show iso-surfaces of transition densities $n_\nu(\mathbf{r)}$ for excitations highlighted in a).}
    \label{fig:fused_transition_densitites}
\end{figure*}

However, the symmetry analysis of $n_{\nu}(\mathbf{r})$ does not explain why $S_5$ contributes only to the N3/N4 (iminic) shake-up spectra. As shown in Figure~\ref{fig:h2tpp_transition_densitites}g, the transition densities of $S_5$ are symmetric with respect to both $\sigma_v^x$ and $\sigma_v^y$, and thus $S_5$ should be observable in both the pyrrolic and iminic satellite spectra. This selectivity can be understood in terms of the second criterion defined above related to the locality of shake-up probabilities: Since the 1s electron density $n_\mathrm{N1s}(\mathbf{r})$ is highly localized around the respective N atom, the Coulomb integral over the product $n_\mathrm{N1s}(\mathbf{r}) n_\nu(\mathbf{r})$ (Eq.\eqref{eq:core_transition_amplitude}) is non-zero only if $n_\nu(\mathbf{r})$ has (symmetric) contributions in the vicinity of that atom. As apparent from Figure~\ref{fig:h2tpp_transition_densitites}g, the transition density of $S_5$ accumulates near N3 and N4, while there is nearly no transition density in the vicinity of N1 and N2. The satellite intensity of $S_5$ is thus practically zero in the N1/N2 spectra. 
The second criterion also explains the large spectral weight of some satellites. For example, the $S_6$ transition has a very high shake-up probability for the iminic sites due to its strong localization at N3 and N4.
\par

In the experiment, the satellite features are not resolved with the same level of detail as in Figure~\ref{fig:h2tpp_transition_densitites}b. Instead, the peaks around 3.0–3.5~eV merge into a single broad feature (gray shaded area in Figure~\ref{fig:h2tpp_transition_densitites}b), and satellites of somewhat lower intensity, such as Q$_x$ or Q$_y$, are not visible. Due to the limited resolution, the satellites associated with N1/N2 and N3/N4 appear essentially the same. However, the differences in satellite features for the individual sites can be much larger, as in the case of the fused porphyrins discussed next, where they are also visible in experiment.

\subsubsection{Fused Porphyrin}
\label{sec:fused}

\begin{table}
    \centering
        \fontsize{10}{12}\selectfont
    \begin{tabular}{ccccccc}
    \toprule \textbf{State}  & $\Omega^\nu$ / eV & $f^\nu $& $\gamma^\nu_\mathrm{N1}$&$\gamma^\nu_\mathrm{N2}$ &$\gamma^\nu_\mathrm{N3}$&$\gamma^\nu_\mathrm{N4}$\\
       \midrule
        $S_1$ & 1.51& 0.024 & 0.022 & 0.001 & 0.024 & 0.001 \\
        $S_4$ & 2.10& 0.123 & 0.002 & 0.031 & 0.048 & 0.005  \\
        \bottomrule
    \end{tabular}
    \caption{Theoretical RPA excitation energies $\Omega^\nu$, dimensionless oscillator strength $f_\nu$ (Eq.~\eqref{eq:oscillator_strength}) compared to the dimensionless satellite intensities $\gamma^\nu_\mathrm{N1s}$ (Eq.~\eqref{eq:gamma_definition}) of all four N~1s core-levels in the double-fused porphyrin \textbf{6}.}
    \label{tab:fused_intensities}
\end
{table}

The satellite spectra for the doubly fused porphyrin \textbf{6} are shown together with the theoretical RPA absorption spectrum in Figure~\ref{fig:fused_transition_densitites}a. For comparison, the broadened satellite spectrum of the larger parent compound \textbf{4} is replotted in gray, corresponding to that shown in Figure~\ref{fig:experiment_comparison}d.\par
Fusing two of the phenyl rings to the porphyrin core 
extends the $\pi$- systems of the macrocycle and drastically alters the optical response: the characteristic Gouterman features with distinct Q- and B-bands disappear and are replaced by a broad absorption band with multiple transitions of comparable intensity. Relative to \textbf{1}, the lowest excitation ($S_1$) is red-shifted to ca.~1.5~eV, while the absorption maximum remains at comparable energies around 2.8 eV.

Ring fusion also drastically changes the satellite spectra, which, as observed for \textbf{1}, differ significantly from the optical spectra. As described in Section~\ref{sec:results_experiment}, the iminic and pyrrolic nitrogen sites are no longer chemically equivalent, leading to distinct spectra for each site.  Compared to \textbf{1}, the satellite spectra of \textbf{6} exhibit more features because no transitions are symmetry-forbidden. Consequently, all excited states can, in principle, contribute, and the main factor determining the shake-up probability is the local weight of the transition density. This is illustrated in Figure~\ref{fig:fused_transition_densitites}b and c for the $S_1$ and $S_4$ excitations.\par

The $S_1$ transition exemplifies why the satellite spectra of \textbf{6} are significantly more complex than those of \textbf{1}. The $S_1$ transition density is fully delocalized across the macrocycle. Since the core-level symmetry is broken, no excited states can be excluded based on symmetry, and the delocalization prohibits locality-based arguments. The interpretation of the $S_1$ satellite becomes thus non-trivial: the low intensity of the $S_1$ satellite in the N2 and N4 spectra (see Table~\ref{tab:fused_intensities}) arises primarily from phase fluctuations across the extended $\pi$-system.
In contrast, the $S_4$ transition (Figure~\ref{fig:fused_transition_densitites}c) is strongly localized on one side of the macrocycle, with density concentrated around the N2/N3 nitrogen sites. As a result, the $S_4$ excitation couples strongly to the corresponding N1s levels, while its interaction with the N1/N4 core levels is negligible.\par

Although the spectra differ at each site, they can still be qualitatively classified into two groups. These groups are not determined by the chemical type of the nitrogen atoms (pyrrolic vs. iminic, N1/N2 vs. N3/N4), but instead by their position relative to the fused rings. The shake-up spectra of the sites on the opposite side of the fused rings (N1 and N4) are similar, showing maxima around 2.8–2.9 eV, which originate from transitions $S_{11}$–$S_{16}$ (see Table S5 in the SI). They also exhibit a pronounced gap between 1.8 and 2.8 eV, in which no transitions carry significant spectral weight. In contrast, the nitrogen sites adjacent to the fused rings (pyrrolic N2 and iminic N3) exhibit maximum intensity within this spectral gap, dominated by the $S_4$ transition at 2.1~eV.
\par
These observations explain why the pyrrolic satellite is no longer observed as a distinct feature after ring fusion. In the doubly fused parent compound \textbf{4}, the broadened theoretical spectra (Figure~\ref{fig:fused_transition_densitites}a, gray) show maximum satellite intensities at 2.95 eV (N1) and 2.15 eV (N2); see also Table~\ref{tab:spectrum_properties}. These values coincide with the intensity maxima of N1 and N2 in the satellite spectra of \textbf{6}, arising from the $S_{16}$ and $S_{4}$ transitions, respectively, with the slight shifts relative to \textbf{4} attributed to the missing tert-butyl groups. In short, N1 and N2 couple most strongly to different charge-neutral excitations separated by 0.8 eV, summing up to a broad tailing feature in the total spectrum rather than a distinct peak.\par

For the mono-fused porphyrin \textbf{5} (see Section~S2 in the SI), the results and observations are qualitatively similar to those of the doubly fused systems, though the changes in the satellite spectrum are less pronounced. 
\section{Conclusion}
\label{sec:conclusion}
We have shown in a combined experimental and theoretical study that ring fusion in porphyrin derivatives induces distinct and chemically relevant changes in the nitrogen 1s shake-up satellite spectra, while the main N~1s ionization peaks remain insensitive to these changes. In particular, we found that the pyrrolic satellite peak progressively vanishes and is replaced by a broad tailing feature as the system evolves from unfused, to mono-fused, to doubly-fused derivatives. This demonstrates that satellite features can be used to monitor structural modifications in XPS, which are not accessible from the photoionization peaks.
\par
The mechanism behind the disappearance of the satellite was clarified by theoretical calculations. We not only decomposed the satellite spectrum into contributions from the individual nitrogen atoms and assigned each feature to its originating core level, but also related the satellites to the underlying charge-neutral valence transitions. This led to the following understanding of why the satellite vanishes: in unfused porphyrins, the pyrrolic nitrogens couple to the same excitations, giving rise to a distinct peak. In fused derivatives, however, the nitrogen core holes couple to different excitations separated by $\approx$ 1~eV, which sum to a broad feature rather than a well-defined satellite peak. We traced these distinct couplings back to locality rules embedded in the shake-up mechanism.
\par
Such calculations have become feasible through our $GW+C$ framework, which we recently developed and further advanced in this work. This fully \textit{ab initio} many-body framework treats both main and satellite features in the spectra on the same footing, yielding spectral functions that can be directly compared with experimental XP spectra. We demonstrated not only excellent agreement between the calculated and experimental spectra, but also, to the best of our knowledge, the largest full ab initio calculations of shake-up satellites reported so far, encompassing systems of up to 170 atoms. Our new computational method has also enabled an understanding of the selection rules governing when satellites appear and why their intensities differ so significantly from those in optical spectra, even though both originate from charge-neutral valence excitations. We showed that the symmetry and locality of the transition density determine the probability of the shake-up weight.
\par
The newly developed computational tool and the insights gained in this work pave the way for extracting additional chemical information from standard XPS measurements. Our future work will focus on improving the description of charge-neutral excitations in $GW+C$ beyond the RPA, incorporating exchange–correlation effects through approaches such as BSE, and by explicitly accounting for core-hole final-state effects. We also aim to extend this framework to core levels beyond~1s by including spin–orbit coupling, as well as to periodic structures.

\begin{acknowledgement}
\fontsize{10}{12}\selectfont

D. G. acknowledges funding by the Emmy Noether Programme of the German Research Foundation (project number 453275048). We also acknowledge funding by the German Research Foundation (GRK2861–491865171).
We gratefully acknowledge the computing time provided on the high-performance computer Noctua 2 at the NHR Center PC2. We thank Prof. Norbert Jux for supporting C. O. and providing access to laboratory facilities that contributed to this work.
\end{acknowledgement}
\begin{suppinfo}
 The supplementary information is available free of charge.

\end{suppinfo}

\bibliography{main}

\end{document}